\documentclass[submission,copyright,creativecommons]{eptcs}
\providecommand{\event}{TTC 2011} 

\usepackage{multirow}
\usepackage{graphicx}
\usepackage{subfigure}
\usepackage{url}
\usepackage{amssymb}
\usepackage{xspace}
\usepackage{hyperref}

\usepackage{amsmath}
\usepackage{listings}

\makeindex

\newcommand{\eg}{e.\,g.\ }

\newcommand{\figref}[2][]{Figure~\ref{#2}#1}

\newcommand{\identifier}[1]{\textsf{#1}}

\newcommand{\myparagraph}[1]{\vspace{1em}\par\noindent\textbf{#1.}}

\lstdefinelanguage{groovy} {
    emph={println, new, tokenize, each, def, static, for, if, else, in, assert},
    emphstyle=\bfseries,
    morecomment=[l]{//},
    basicstyle=\fontfamily{pcr}\scriptsize,
    string=[b]",
    showstringspaces=false
}

\title{GMF: A Model Migration Case for the Transformation Tool Contest}
\author{Markus Herrmannsdoerfer
\institute{Institut f\"ur Informatik,
  Technische Universit\"{a}t M\"{u}nchen 
}
\email{herrmama@in.tum.de}
}
\def\titlerunning{GMF: A Model Migration Case for the Transformation Tool Contest}
\def\authorrunning{M. Herrmannsdoerfer}
\begin{document}
\maketitle

\begin{abstract}
Using a real-life evolution taken from the Graphical Modeling Framework, we invite submissions to explore ways in which model transformation and migration tools can be used to migrate models in response to metamodel adaptation.
\end{abstract}

\section{Model Migration}

Modeling languages and thus their metamodels are subject to evolution~\cite{Favre2003_Meta-ModelandModelCo-evolutionwithinthe3DSoftwareSpace}.
When a metamodel is adapted, existing models may no longer conform to the adapted metamodel and thus need to be migrated.
Model migration is a special case of exogenous model transformation~\cite{Mens2006_ATaxonomyofModelTransformation}, since original and adapted metamodel are usually different from each other.
However, the metamodel versions also share some similarity, as the metamodel is usually not completely changed during metamodel adaptation~\cite{Sprinkle2004_ADomain-SpecificVisualLanguageForDomainModelEvolution}.
Consequently, migrating transformation definitions usually contain identity rules for the unchanged metamodel parts. 
To remove this boilerplate code, different approaches have been proposed~\cite{Rose2009_AnAnalysisofApproachestoModelMigration}.

\emph{Manual specification} approaches---like Sprinkle's language~\cite{Sprinkle2004_ADomain-SpecificVisualLanguageForDomainModelEvolution}, MCL~\cite{Narayanan2009_AutomaticDomainModelMigrationtoManageMetamodelEvolution} and Epsilon Flock~\cite{Rose2010_ModelMigrationwithEpsilonFlock}---extend transformation languages so that they automatically copy model elements that are unaffected by metamodel adaptations.
\emph{Operator-based} approaches---like Ecoral~\cite{Wachsmuth2007_MetamodelAdaptationandModelCo-adaptation} and COPE~\cite{Herrmannsdoerfer2009_COPE-AutomatingCoupledEvolutionofMetamodelsandModels}---provide reusable operators that encapsulate recurring metamodel adaptations and model migrations.
\emph{Metamodel matching} approaches---like Cicchetti's approach~\cite{Cicchetti2008_AutomatingCo-evolutioninModel-DrivenEngineering} and AML~\cite{Garces2009_ManagingModelAdaptationbyPreciseDetectionofMetamodelChanges}---automatically derive a transformation definition from the difference between two metamodel versions.
The existing approaches mostly use or extend existing model transformation languages and tools.

To compare the different ways in which model migration can be defined, we propose a real-life case from the evolution of the Graphical Modeling Framework (GMF).
The case is already well-researched, as it has been used in an empirical~\cite{Herrmannsdoerfer2010_LanguageEvolutioninPracticeTheHistoryofGMF} and a comparative study~\cite{Rose2010_AComparisonofModelMigrationTools}.
It exhibits a number of differences to last year's migration case~\cite{Rose2010_ModelMigrationCaseforTTC2010}:
(1)~Most of the metamodel remains the same which is more typical for model migration.
(2)~The migration is more complex and thus requires more expressive transformation languages.
(3)~GMF provides a reference migrator implemented in Java which defines the migration semantics.
(4)~GMF provides a number of test cases which can be used for validating the solutions.

\section{Graphical Modeling Framework}

\begin{figure}[tb]
 	\centering
  \subfigure[Definition]{
    \includegraphics[scale=.5]{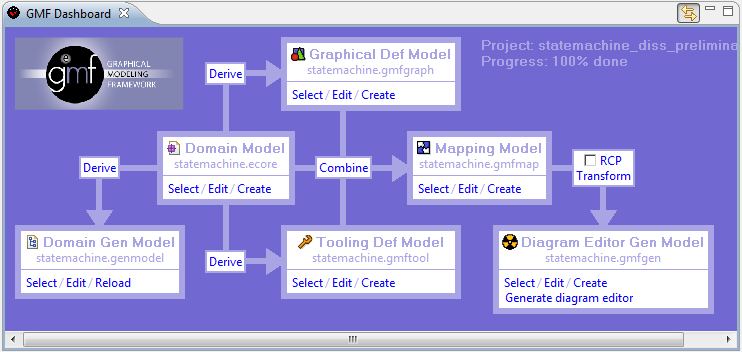}
    \label{fig:preliminaries_gmf_definition}
  }
  \subfigure[Implementation]{
		\includegraphics[scale=.5]{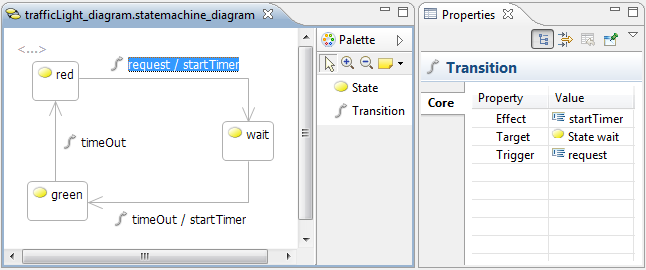}
    \label{fig:preliminaries_emf_concrete_syntax_diagrammatic}
  }
		\vskip -5pt
  \caption{Graphical Modeling Framework}
  \label{fig:preliminaries_emf_concrete_syntax_diagrammatic_main}
\end{figure}

The Graphical Modeling Framework (GMF)\footnote{\url{http://www.eclipse.org/modeling/gmp/}} is a widely used open source framework for the model-driven development of diagram editors based on the Eclipse Modeling Framework (EMF).
GMF is a prime example for a Model-Driven Architecture (MDA) \cite{Kleppe2003_MDAExplainedTheModelDrivenArchitecturePracticeandPromise}, as it strictly separates platform-independent models (PIM), platform-specific models (PSM) and code.
GMF is implemented on top of the Eclipse Modeling Framework (EMF)\footnote{\url{http://www.eclipse.org/modeling/emf}} and the Graphical Editing Framework (GEF)\footnote{\url{http://www.eclipse.org/gef}}.
\figref{fig:preliminaries_gmf_definition} shows the GMF dashboard which supports the process of creating the models of the diagram editor.

The GMF models are based on a \identifier{domain model} expressed in Ecore from which an appropriate \identifier{domain generator model} can be derived.
GMF provides wizards to derive the following models from the domain model:
the \identifier{graphical definition model} defines the graphical elements like nodes and edges in the diagram, and the \identifier{tooling definition model} defines the tools available to author a diagram.
The \identifier{mapping model} maps
graphical elements from the \identifier{graphical definition model} and the tools from the \identifier{tooling definition model} to the constructs from the \identifier{domain model}.
The \identifier{mapping model} is transformed into a \identifier{diagram generator model} from which a diagram editor can be generated. 
The \identifier{diagram generator model} can be altered to customize the code generation.

\figref{fig:preliminaries_emf_concrete_syntax_diagrammatic} shows the diagram editor generated from GMF models defined for a statemachine modeling language.
The diagram editor shows the graphical elements in the diagram and the tools in the palette.
GMF also provides more advanced features like \eg annotating, zooming and layouting for the generated editor.
The properties of a graphical element can be accessed through the properties view.

\section{Evolution of GMF Graph}
\label{sec:gmfgraph}

Here, we consider one of those metamodels---GMF Graph---to which graphical definition models have to conform.

\begin{figure}[tb]
 	\centering
  \subfigure[Metamodel]{
    \includegraphics[scale=.6]{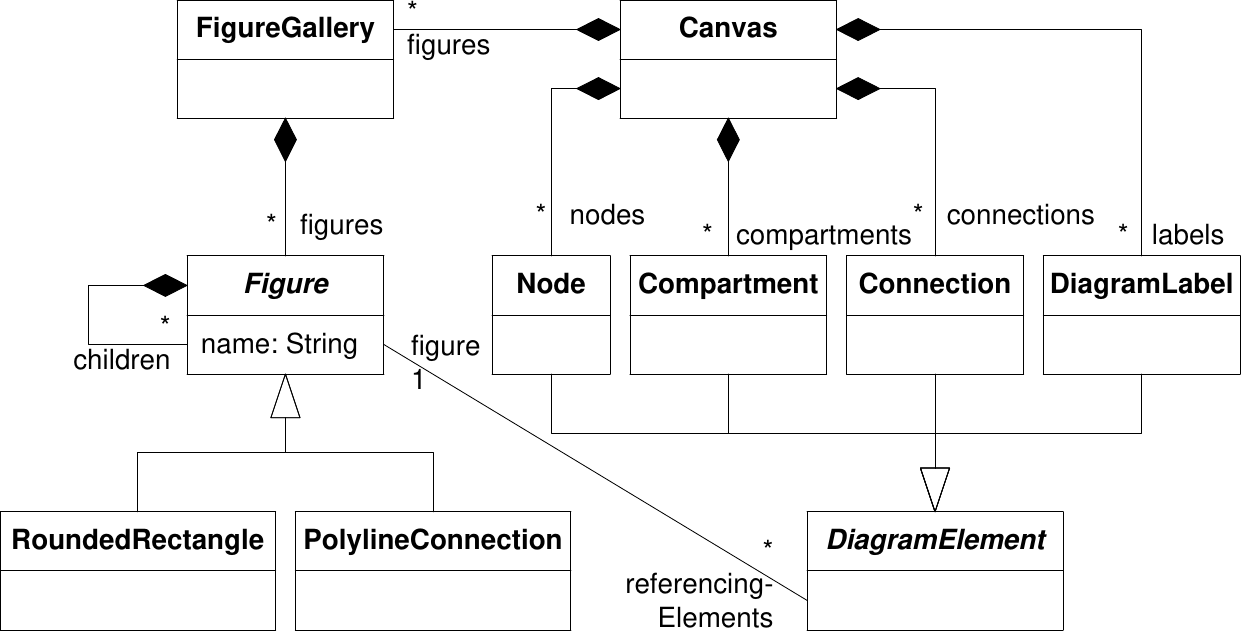}
    \label{fig:gmfgraph1_metamodel}
  }
  \qquad
  \subfigure[Model]{
		\includegraphics[scale=.5]{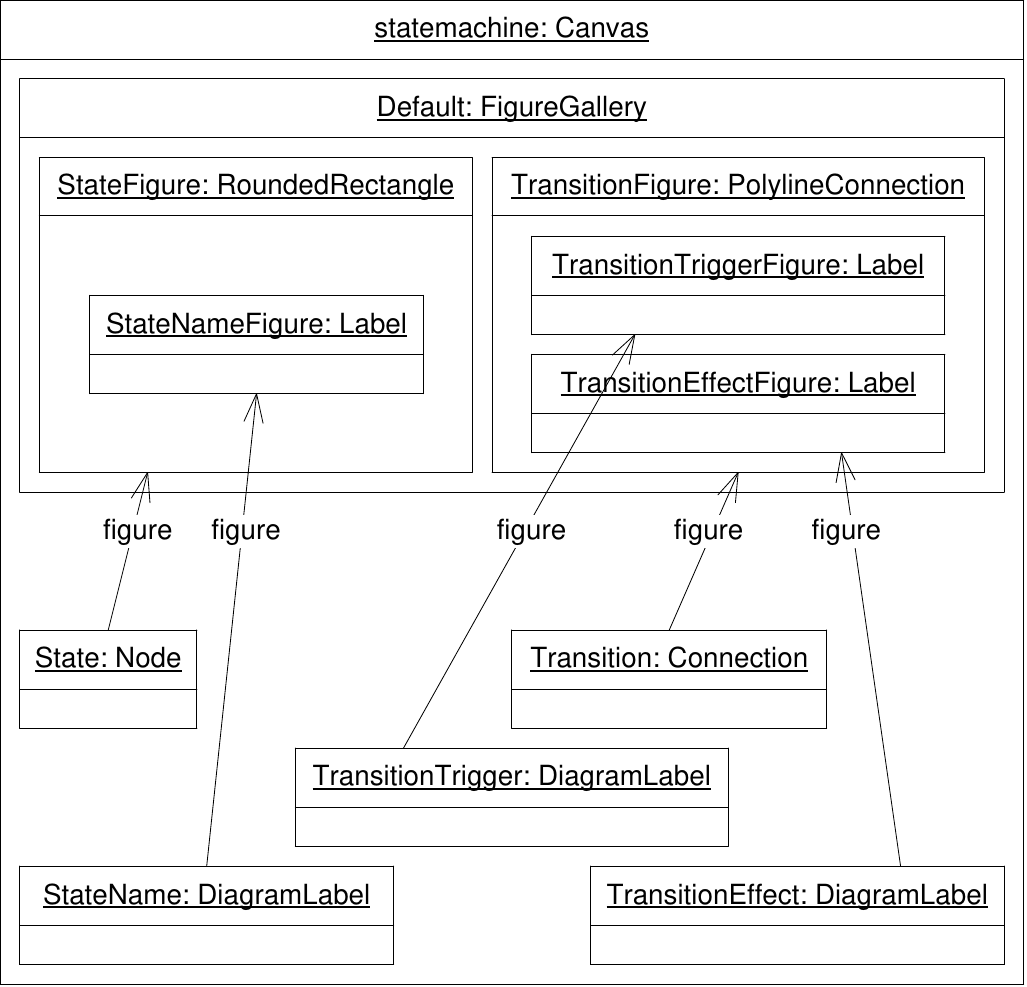}
    \label{fig:gmfgraph1_model}
  }
		\vskip -5pt
  \caption{GMF version 1.0}
  \label{fig:gmfgraph1}
\end{figure}

Figure~\ref{fig:gmfgraph1_metamodel} shows the part of the GMF Graph metamodel that has changed from GMF version 1.0 to 2.1 as a class diagram.
The GMF Graph metamodel describes the appearance of the generated diagram editor. The classes \identifier{Ca\-nv\-as}, \identifier{Fi\-gu\-re}, \identifier{No\-de}, \identifier{Di\-ag\-r\-amLa\-b\-el}, \identifier{Co\-nn\-ec\-ti\-on}, and \identifier{Co\-mp\-ar\-tm\-e\-nt} are used to represent components of the diagram editor to be generated. 
Figure~\ref{fig:gmfgraph1_model} shows the model to define the graphical elements of the statemachine modeling language in a modified object diagram---objects that are contained in other objects  by means of a composition are shown inside these objects.

The evolution in the GMF Graph metamodel was driven by analyzing the usage of the \identifier{Fi\-gu\-re.re\-fe\-re\-nc\-ing\-El\-em\-en\-ts} reference, which relates \identifier{Fi\-gu\-res} to the \identifier{Dia\-gram\-Ele\-ments} that use them. As described in the GMF Graph documentation\footnote{\url{http://wiki.eclipse.org/GMFGraph_Hints}}, the \identifier{re\-fe\-re\-nc\-ing\-El\-em\-en\-ts} reference increased the effort required to reuse figures---a common activity for users of GMF. Furthermore, \identifier{re\-fe\-re\-nc\-ing\-El\-em\-en\-ts} was used only by the GMF code generator to determine whether an accessor
should be generated for nested \identifier{Fi\-gu\-re}s.

\begin{figure}[!tb]
 	\centering
  \subfigure[Metamodel]{
    \includegraphics[scale=.6]{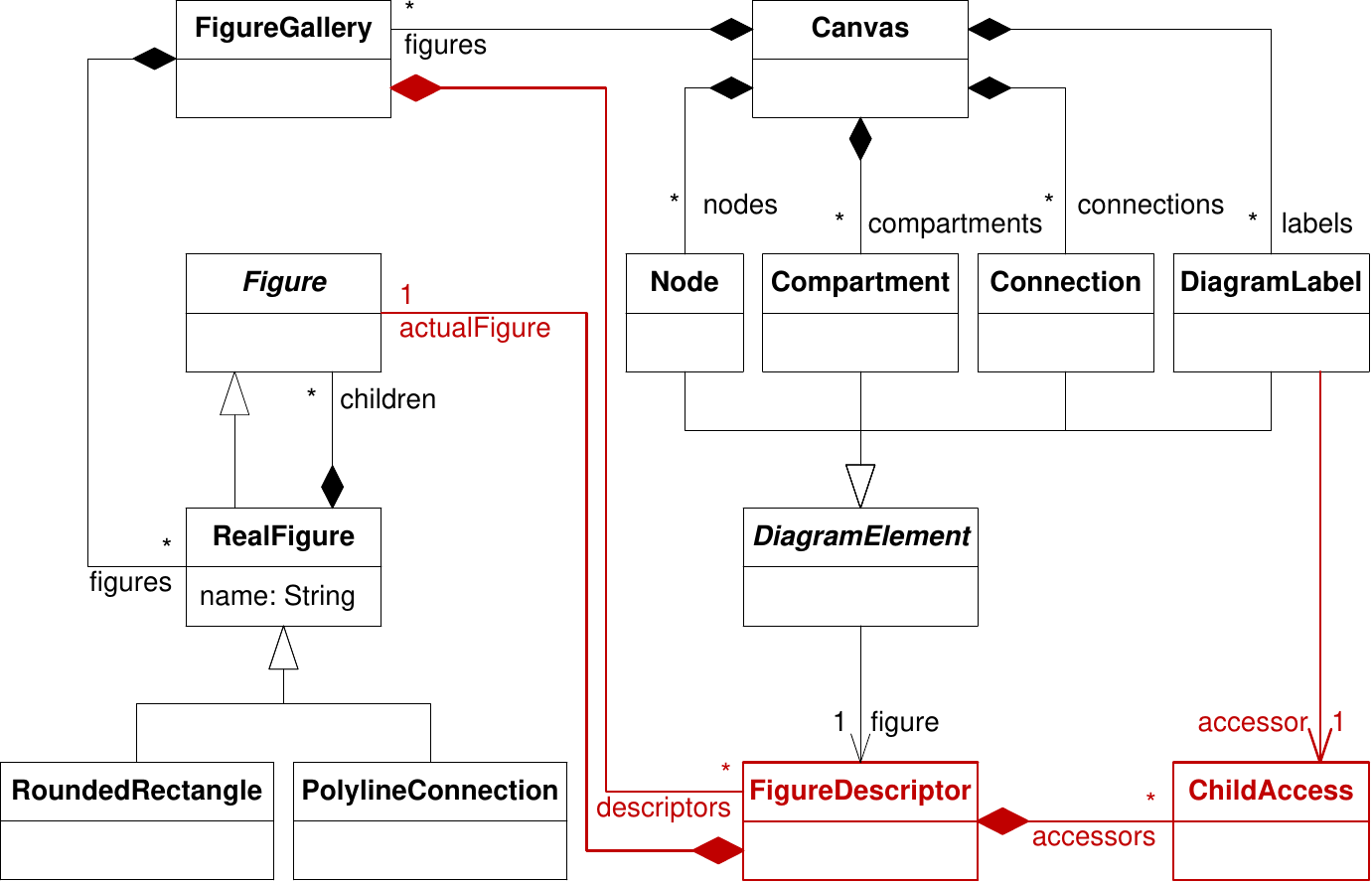}
    \label{fig:gmfgraph2_metamodel}
  }
  \hfill
  \subfigure[Model]{
		\includegraphics[scale=.5]{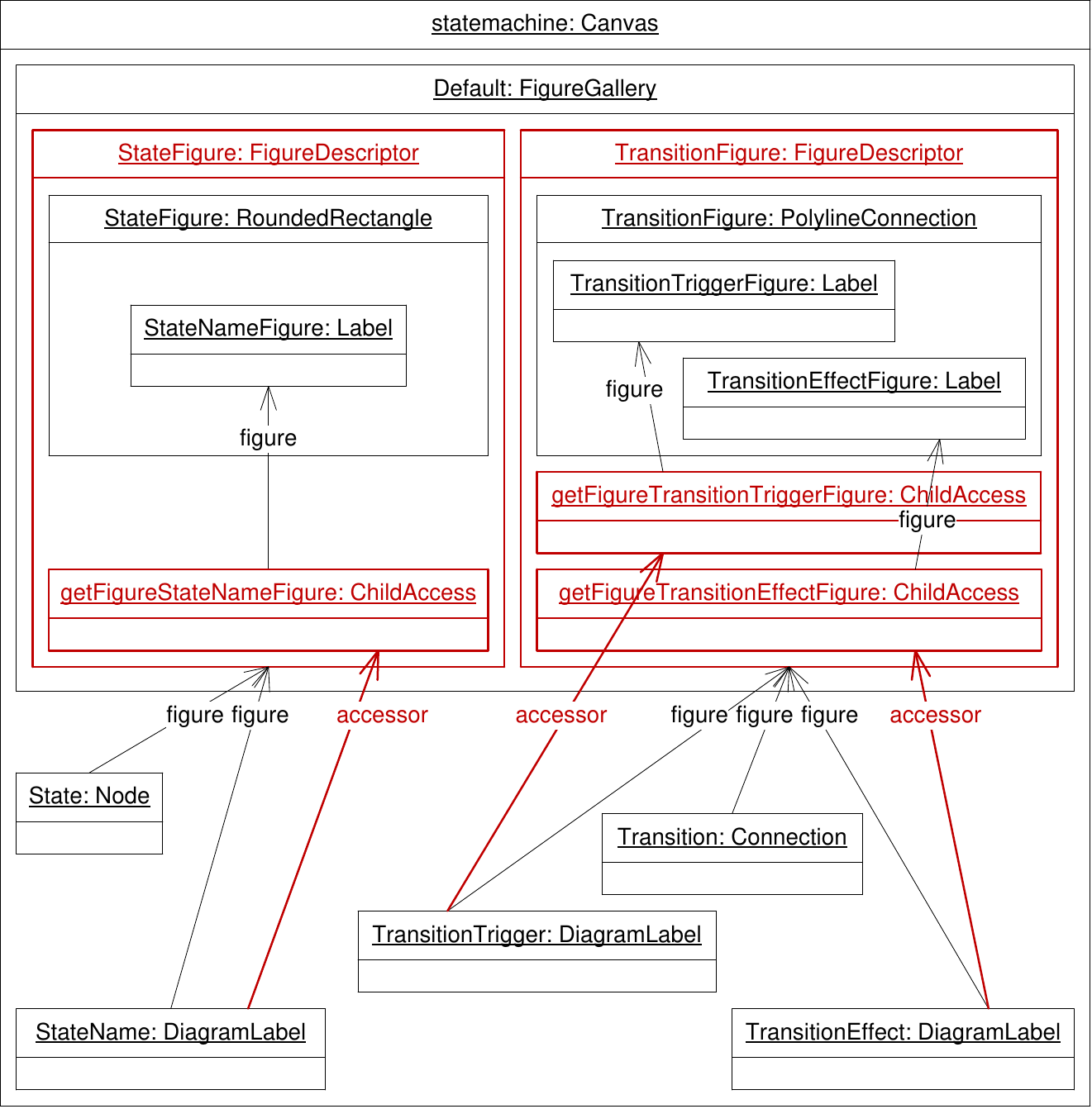}
    \label{fig:gmfgraph2_model}
  }
		\vskip -5pt
  \caption{GMF version 2.1}
  \label{fig:gmfgraph2}
\end{figure}

In GMF 2.1, the Graph metamodel was evolved to make reusing figures more straightforward by introducing a proxy for \identifier{Fi\-gu\-re}, termed \identifier{Fi\-gu\-reDe\-scrip\-tor}. Figure~\ref{fig:gmfgraph2_metamodel} highlights the changes that have been made to the GMF Graph metamodel. The original \identifier{re\-fe\-re\-nc\-ing\-El\-em\-en\-ts} reference was removed, and an extra class, \identifier{Ch\-ild\-Ac\-ce\-ss}, was added to make more explicit the original purpose of \identifier{re\-fe\-re\-nc\-ing\-El\-em\-en\-ts}---accessing nested \identifier{Fi\-gu\-re}s.
Figure~\ref{fig:gmfgraph2_model} shows the migrated model of the graphical elements of the statemachine modeling language.
Instances of \identifier{FigureDescriptor} and \identifier{ChildAccess} have been introduced as interfaces for the figures that can now be more easily reused using these interfaces.

GMF provides a migrator that produces a model conforming to the evolved Graph metamodel from a model conforming to the original Graph metamodel\footnote{\url{http://wiki.eclipse.org/GMF_Migration}}.
In GMF, the migration is implemented in Java\footnote{\url{http://dev.eclipse.org/viewcvs/viewvc.cgi/org.eclipse.gmp/org.eclipse.gmf.tooling/plugins/org.eclipse.gmf.graphdef/src/org/eclipse/gmf/internal/graphdef/util/MigrateFactory2005.java?root=Modeling_Project&view=log}}. 
The version control system of GMF provides test models to validate the migrator which can be used to validate the submitted solutions\footnote{\url{http://dev.eclipse.org/viewcvs/viewvc.cgi/org.eclipse.gmp/org.eclipse.gmf.tooling/tests/org.eclipse.gmf.tests/models/migration/?root=Modeling_Project}}.
The GMF source code includes two example editors, for which the version control system contains versions conforming to GMF 1.0 and GMF 2.1. 

\section{Tasks and Criteria}

\myparagraph{Core Task}
The case consists of a core task and two extensions.
The core task is to use a model transformation or migration tool to perform the migration explained in Section~\ref{sec:gmfgraph}.
Submissions are evaluated according to the following criteria:

\begin{itemize}
	\item
	\textbf{Expressiveness}:
	Is the language on which the transformation tool is based expressive enough to specify the migration?
	Or is it required to use a programming language to be able to completely specify the migration?

	\item 
	\textbf{Correctness}:
	Does the transformation correctly migrate the test cases included in the case resources?
	Does the transformation produce a model equivalent to the model migrated by the GMF migrator?

\item 
	\textbf{Conciseness}:
	How much code is required to specify the transformation?
	Is the size of the code proportional to the number of differences between the metamodel versions or to the size of the metamodel versions?

\item 
	\textbf{Maintainability}:
	How easy is it to read and understand the transformation?
	How easy is it to modify or extend the transformation?
\end{itemize}

The case resources include the different GMF Graph metamodel versions as well as the test models.

\myparagraph{Multi-File Models}
In GMF, models can be split over several files to be able to modularize them.
When migrating a multi-file model, the modularization should be preserved by the transformation.
In this extension, submissions should show that they are able to preserve the modularization into files.
To facilitate this, the case resources include multi-file models for both metamodel versions.

\myparagraph{GMF Map metamodel}
The GMF Map metamodel defines the structure to which mapping models need to conform.
It has also been adapted from GMF 1.0 to GMF 2.1, but the migration is much more simple than in case of GMF Graph\footnote{\url{http://dev.eclipse.org/viewcvs/viewvc.cgi/org.eclipse.gmp/org.eclipse.gmf.tooling/plugins/org.eclipse.gmf.map/src/org/eclipse/gmf/internal/map/util/?root=Modeling_Project}}.
However, this migrating transformation poses a number of other challenges.
(1) The GMF Map metamodel refers to other metamodels, and thus GMF Map models also refer to models of these metamodels.
A transformation needs to be able to preserve these references.
(2) The GMF Map metamodel has been released one more time than the GMF Graph metamodel.
Thus, a transformation needs to detect the version of the model and appropriately migrate it to the newest version of the metamodel.
To be able to do this, the transformation tool needs to be able to chain transformation definitions~\cite{Pilgrim2008_ConstructingandVisualizingTransformationChains}.
In this extension, submissions should show that they are able to cope with these challenges.
To facilitate this, the case resources include the different versions of the GMF Map metamodel as well as test models.


\bibliographystyle{eptcs}
\bibliography{ttc2011}

\appendix

\end{document}